\documentclass[aps,prd,twocolumn,superscriptaddress]{revtex4}
\usepackage{graphicx}% Include figure files
\usepackage{amsmath}
\usepackage{amssymb}
\usepackage{slashed}
\usepackage{latexsym}
\usepackage{epsfig}
\usepackage{amsbsy}
\usepackage{array}
\usepackage{amssymb}
\usepackage{changes}
\usepackage{color}
\usepackage{setspace}
\usepackage{bm}
\usepackage{lipsum}
\usepackage{mathrsfs}
\usepackage{float}
\usepackage{color}
\usepackage[T1]{fontenc}
\usepackage{mathptmx}
\DeclareMathAlphabet{\mathcal}{OMS}{cmsy}{m}{n}
\DeclareSymbolFont{largesymbols}{OMX}{cmex}{m}{n}

\begin{document}

\title{  Non-strange quark stars from NJL model with proper-time regularization}

\author{Qingwu Wang }
 \email{qw.wang@scu.edu.cn }
 \affiliation{College of physics, Sichuan University, Chengdu 610064, China}

  \author{Chao Shi}
  \email{ shichao0820@gmail.com}
  \affiliation{College of Material Science and Technology,
Nanjing University of Aeronautics and Astronautics, Nanjing 210016, China;}

\author{Hong-Shi Zong}
\email{zonghs@nju.edu.cn}
\affiliation{Department of Physics, Nanjing University, Nanjing 210093, China}
\affiliation{Joint Center for Particle, Nuclear Physics and Cosmology, Nanjing 210093, China}

\begin{abstract}

The structure of light quark star   is studied within a new two-flavor NJL model. By retaining the contribution from the vector term in the  Fierz-transformed Lagrangian, a two-solar-mass pure quark star is achieved. To overcome the disadvantage of three-momentum truncation in the regularization  procedure, we introduce the proper-time regularization.  We also employ the newly proposed definition of vacuum pressure  \cite{xu2018}, in which the quasi-Wigner vacuum (corresponding to the quasi-Wigner solution of the gap equation) is used as the reference ground state.
Free parameter  includes only  a mixing constant $\alpha$ which  weighs   contribution  from Fierz-transformed Lagrangian.  We constrain $\alpha$ to be around $0.9$   by the observed  mass of pulsars $PSR J0348+0432$ and $PSR J1614-2230$. We find  the calculated surface energy density meets the requirement ($> 2.80\times10^{14}$g/cm$^3 $) \cite{libl2019}. Besides, for a 1.4 solar mass star, the   deformability $\Lambda$ is calculated which is consistent with   a recent analysis on the binary neutron star merger GW170817 with $\Lambda$ in (0, 630) for large component spins and 300$^{+420}_{-230}$  when restricting the magnitude of the component spins \cite{Abbott2019}, and satisfies the constraints  $200 < \Lambda < 800$ of early works \cite{Bauswein2019,Margali,Abbott}

%\pacs{genera qcd  12.38.Aw, bag 12.39.Ba, quarks 14.65.Bt, nstar97.60.Jd}

\end{abstract}

\maketitle
\section{Introduction}
Investigations of  the dense matter  is   an important part of studying strong interactions. The experimental data on the ground tells us that the ground state of the strong interacting   baryon at  zero density is nucleon.
At non-zero density, when the quark chemical potential is higher than the strange quark mass, the strange quark matter may be the ground state \cite{Witten} . Therefore, the observed pulsars may be   quark star rather than a traditional neutron star. After the discovery of two solar-mass pulsars  \cite{Demorest,Antoniadis,Fonseca,Cromartie}, many theoretical models about non-strange star were excluded because  they lead to equations of state (EOSs)   that were too  soft. Considering the   EOS of strange baryon (hyperon), the maximum pulsar mass is   still lower than 2 $M_{\odot}$.    But the structure   of neutron stars can  be   explored   by considering  modified theory of  gravity  \cite{DOBADO, BABICHEV}, introducing a strong electromagnetic field \cite{RAY,GHEZZI,PICANC}, or introducing a high-speed rotation effect  \cite{COOK,STERGIOULAS,KRASTEV} to obtain a 2 $M_{\odot}$ neutron star.
  Recently one literature indicates that a stable hadron matter can be a non-strange quark state when the  baryon number is greater than a certain minimum value \cite{Holdom}. Quark matter with only $u$-$d$ quarks can be ground state of baryonic matter other than the $u$-$d$-$s$ strange quark matter \cite{Witten,Bodmer,Terazawa}. Therefore, if the observed pulsar is a non-strange quark star and regardless of all other corrections,    a suitable   EOS becomes necessary.

Studying the   EOS at extreme environment usually   resorts to   effective field theories, and  the Nambu$-$Jona-Lasinio (NJL) model is one of them.  With a few parameters   fitted to the low energy experimental data, the NJL  model and its many generalizations provided much information on both hadron physics and QCD matter at finite temperature and density \cite{Buballa,Klevansky}. We note within the NJL model, there are two equivalent descriptions, i.e., the original Lagrangian density $L_{NJL}$   and its  Fierz transformation $L_{Fierz}$.  After doing the mean field approximation, $\left<L_{NJL}\right>$ contains only the Hartree term, while $\left<L_{Fierz}\right>$ contains only the Fock term. When calculating the quark condensation, the parameters are calibrated to   reproduce the physical pion meson mass and decay constant. In this way, the two descriptions are   equivalent.
However, when considering finite density   matter, the effective chemical potentials given by the two   Lagrangians are different at mean field approximation level: the effective potential from the Fierz transformed Lagrangian  has contribution from quark vector density $\left<\psi^\dagger\psi\right>$.   Actually, without the Fierz transformation, the standard approach of mean field approximation  is   considered  to be not self-consistent theoretically \cite{Kunihiro}. We therefore propose a self-consistent   treatment that    combines the  $\left<L_{NJL}\right>$ and $\left<L_{Fierz}\right>$ linearly with a weighting parameter $\alpha$ and study its implication on QCD matter at finite density  \cite{wangf}.  Based on this, our previous study  \cite{zhaot,wangqy} has shown that two-flavor
quark matter could be more stable than the three-flavor quark matter, unlike Witten's prediction based on MIT bag model   \cite{Witten}.  As stated in the Ref. \cite{wangf}, the parameter $\alpha$ used to reflect the weight of different interaction channels can not be given in advance by the mean field theory. It must be determined by related experimental data of high density strong interacting matter. The neutron star provides such a laboratory. One motivation of this paper is to determine the $\alpha$ through current astronomical observations on neutron stars.

In this paper, we will extend our study in several new aspects. We first employ the proper-time regularization  rather than the three-momentum cutoff scheme, since the latter   limits the value of $\alpha<0.9$ and hence considered to be less reliable  \cite{zhaot}. Note that for a non-renormalizable theory, regularization scheme could play an important role in making physical predictions. Testing with different  regularization  schemes therefore provides a qualitative check for consistency.
 Secondly, when calculating the  EOS, there is a free parameter , i.e., the bag constant $B$.
 The bag constant $B$ gives the pressure of quark matter at zero temperature and zero density.  Usually, it is treated as a phenomenological parameter and determined by experimental requirements \cite{licm,wangqy,zhaot,Chodos,Alcock,yany}.
It has a great influence on the    EOS and consequently the mass-radius relation of the neutron star. We should be more careful in choosing    its value.    Experimental requirements  \cite{Alford,Zhou,libl2019} suggest a  typical value about ($120$ MeV)$^4$, but it's also calculable in effective theories.
A  traditional treatment is to subtract the thermodynamic potential of the current quark from the thermodynamic potential corresponding to the Nambu-Goldstone solution, i.e., $B=\Omega(m_{current})-\Omega(M_{Nambu})$ \cite{Buballa}. But the current quark is not the solution to the gap equation. We therefore use a recently proposed definition $B=\Omega(M_{Wigner})-\Omega(M_{Nambu})$ \cite{licm,xu2018,cui2018}, i.e., to subtract from the thermodynamic potential corresponding to the Wigner-Weyl solution.   Such a definition is more theoretically self-consistent since the Wigner-Weyl solution is another (although unphysical) solution to the quark gap equation in NJL model. The bag constant is now the pressure difference between the Nambu and Wigner phases. It is an intrinsic quantity within the NJL model, instead of an input parameter as was in  Ref. \cite{zhaot}. Finally, we determine  the   weighting factor $\alpha$ from the   experimental data. Besides the observations of the pulsars mass, the tidal deformability measurement  from the neutron star merger $GW170817$  \cite{Abbott2019,Abbott,Gao,Bauswein2017,Christian,Bauswein2019}  has also been used to constrain the stiffness of the   EOS. We will check if   our predictions agree with other current observations on, e.g., radius and surface energy of star.
%%%%%%%%%%%%%%%%%%%%%%%%%

This paper is organized as follows: In Sec. II, we introduce  the NJL model and its Fierz transformation. The   weighting factor $\alpha$ is introduced here and its effects on the quark mass and quark number density are presented.   In Sec. III, The   EOS, the mass-radius relation  and tidal deformability of the stars are calculated. The effect of $\alpha$ on the   EOS is  presented. In the end, a short   summary is given.

\section{ the NJL model and its Fierz transformation}
The standard NJL model is a   model of QCD with four quark interaction  \cite{Buballa,Klevansky}.    Beyond the chiral limit,
  the two-flavor Lagrangian is

  \begin{eqnarray}
\mathcal{L}_{NJL}=\bar{\psi }(i\slashed{\partial } - m)\psi + G[\left(\bar{\psi }\psi\right)^2+\left(\bar{\psi } i \gamma ^5 \vec{\tau }\psi \right)^2]+\mu  {\psi^\dagger }\psi.
\end{eqnarray}
  In the mean field approximation,
   \begin{eqnarray}
\mathcal{L}_1=\bar{\psi }(i\slashed{\partial } - m)\psi + 2 G \sigma_1 \bar{\psi }\psi +\mu  {\psi^\dagger }\psi   ,
\end{eqnarray}
where $m$ is the current quark mass and $G$ is the four-quark effective coupling. The two-quark condensate is denoted as $\sigma_1$.
 The effective quark mass is defined as
 \begin{equation}\label{eq.m1}
 M=m-2G\sigma_1,
 \end{equation}
  with the two-quark condensate   defined as
 \begin{equation}\label{eq.gap1}
 \sigma_1=\left\langle {  \bar \psi \psi} \right\rangle=-\int \frac{d^4p}{(2\pi)^4}\textrm{Tr}[S(p)],
\end{equation}
where $S(p)$ is the dressed quark propagator and the trace is taken in color, flavor and Dirac spaces.
 The integration is divergent and a cut-off $\Lambda$ on the momentum is   usually used.   In this case, the chemical potential $\mu$ must be less than the
 cut-off $\Lambda$ so as to get a reliable result, setting an   upper limit for the weighting parameter $\alpha$ \cite{zhaot}.
 To circumvent this defect, we introduce the proper-time regularization here.   The key equation is a replacement
  \begin{equation}
  \frac{1}{A(p^2)^n} \rightarrow \frac{1}{(n-1) !}\int_{\tau_{UV}}^ \infty d\tau \tau^{n-1}e^{-\tau A(p^2)},
  \end{equation}
where $\tau_{UV}$ is introduced to regularize the ultra-violet  divergence.

 %%%%%%%%%%

 Beyond the chiral limit, three parameters ($\tau_{UV}$, $G$, $m$) need to be fixed by requirements such as two-quark condensate derived from QCD sum rules or lattice QCD, pion decay constant or pion mass which are the results of
 chiral symmetry breaking.  Since  the  Gell--Mann$-$Oakes$-$Renner (GMOR) relation is satisfied in low energy,
it also can be used   to calibrate these parameters.
Fixing $m = 5.0$ MeV,  with $G=3.086 \times 10^{-6} $MeV$^{-2}$,
  and $\tau_{UV} = 1092 $ MeV,  it gives $f_\pi$$ = 93 $ MeV  and  $m_\pi$ $= 135$ MeV.  At zero temperature, the condensate is

  \begin{eqnarray}\label{eq.gapt0}
\left\langle {  \bar \psi \psi} \right\rangle&=&-2N_c\sum_{u,d}\int \frac{d^3p}{(2\pi)^3} \frac{M}{E_p}(1-\theta(\mu-E_p))\\
&=&-2N_c\sum_{u,d}\left (\int\frac{d^3p}{(2\pi)^3} \int_{\tau_{ UV}}^\infty d\tau  \frac{e^{-\tau E^2}}{ \sqrt{\pi \tau}} \right. \nonumber\\
 &&\left. -\int \frac{d^3p}{(2\pi)^3} \frac{M}{E_p}\theta(\mu-E_p)\right), \nonumber
\end{eqnarray}
  with $E_p=\sqrt{\vec{p}^2+M^2}$.

 %%%%%%%%%%%%%%
  As a purely technical device to examine the effect of a rearrangement of fermion field operators,  the Fierz transformation of $L_{NJL}$ is:
\begin{eqnarray}
\begin{aligned}
\mathcal{L} _{Fierz}=& \bar{\psi }(i\slashed{\partial } - m)\psi + \frac{G } {8 N_c}[2\left(\bar{\psi } \psi \right)^2+2\left(\bar{\psi } i \gamma ^5 \tau \psi \right)^2\\
& - 2\left(\bar{\psi } \tau \psi \right)^2 - 2\left(\bar{\psi } i \gamma ^5 \psi \right)^2 - 4\left(\bar{\psi } \gamma^{\mu } \psi \right)^2\\
& - 4\left(\bar{\psi } \gamma^{\mu } \gamma^5\psi \right)^2 + \left(\bar{\psi } \sigma^{\mu \nu} \psi \right)^2 - \left(\bar{\psi } \sigma^{\mu \nu} \tau \psi \right)^2].
\end{aligned}
\end{eqnarray}
And the mean-field-approximation result is

\begin{eqnarray}
\mathcal{L} _{2}&=& \bar{\psi }(i\slashed{\partial } - m)\psi +  \frac{G} {2 N_c} \sigma_1 \bar{\psi } \psi +G\sigma_1^2 \nonumber \\
&&+\mu  {\psi^\dagger }\psi- \frac{G} {  N_c}\sigma_2{\psi^\dagger } \psi +\frac{G} { 2 N_c}\sigma_2^2,
\end{eqnarray}
 with $\sigma_2=\left\langle{\psi ^\dagger } \psi\right\rangle$.
 Apparently, the  effective  quark mass and chemical potential in the Fierz-transformed Lagrangian can be defined as:
 \begin{eqnarray}\label{eq.m2}
 M&=&m-\frac{G}{2N_c}\sigma_1,
  \end{eqnarray}
   \begin{eqnarray}\label{eq.mu}
 \mu_r&=&\mu- \frac{G} {  N_c}\sigma_2.
  \end{eqnarray}

  Although the two formulas Eq. (\ref{eq.m1}) and Eq. (\ref{eq.m2})  look different, in fact, when the parameters are recalibrated with different values of coupling $G$, they get the same result of quark mass. If the NJL Lagrangian and its Fierz transformation are combined according to the literature \cite{Klevansky}, the second term  in Eq. (\ref{eq.m2}) is only equivalent to the next-to-leading-order term of large $N_c$ expansion. And the amending in the effective chemical potential almost can also be neglected.

 However, the equivalence of $L_{NJL}$  and $L_{Fierz}$ means   their linear combination with any complex $\alpha$, and in the mean field approximation it reads  :

 \begin{eqnarray}\label{eq.la}
\mathcal{L} _C=\left(1-\alpha \right)\mathcal{L}_1 + \alpha \mathcal{L} _2.
\end{eqnarray}
With this combination,  the effective mass and chemical potential are:

  \begin{eqnarray}\label{eq.mc}
 M&=&m-2(1-\alpha+\frac{\alpha}{4N_c})G\sigma_1,
  \end{eqnarray}
   \begin{eqnarray}\label{eq.muc}
 \mu_r&=&\mu- \frac{\alpha G} {  N_c}\sigma_2.
  \end{eqnarray}
  Since the coefficient $(1-\alpha+\frac{\alpha}{4N_c})G$ in Eq. (\ref{eq.mc}) requires fitting low energy data, we can redefine the effective coupling as
  \begin{equation}
  G^\prime=(1-\alpha+\frac{\alpha}{4N_c})G .
  \end{equation}
Thus, the quark mass and    chemical potential are rewritten as
   \begin{eqnarray}\label{eq.muc2}
    M&=&m-2G^\prime\sigma_1,\\
 \mu_r&=&\mu- \frac{G^\prime} {  N_c}\frac{\alpha }{1-\alpha+\frac{\alpha}{4N_c}}\sigma_2.
  \end{eqnarray}
Only for $\alpha<1$, this is mathematically equivalent to adding a vector--isoscalar channel   in the $L_{NJL}$\cite{Benic}.     By introducing a positive value of $\alpha$  one counts in more terms that may be ignored in  the mean field approximation with only $L_{NJL}$. So, in this   sense, the two approaches are different.

 \begin{figure}
   \begin{center}
      \includegraphics[width=0.45\textwidth]{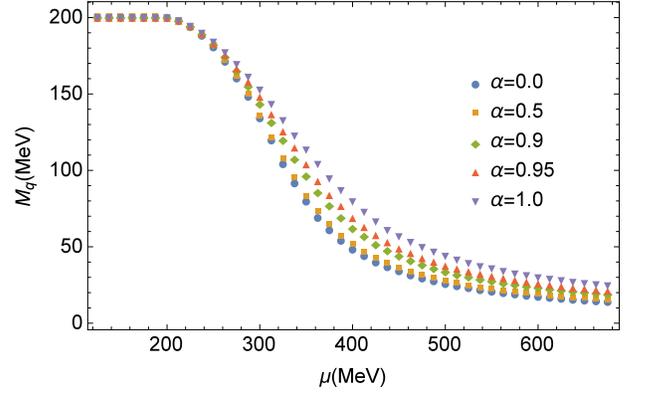}
      \caption{The quark mass as a function of $\mu$ is presented. It is the standard NJL model result with $\alpha=0$, and the Fierz--transformed result with $\alpha=1$. With $\alpha=0.5$, it is the result of taking into account the   next-to-leading order in large $N_c$ expansion which has little impact on effective quark mass.}
\label{fig.mq}   \end{center}
\end{figure}

 \begin{figure}
   \begin{center}
      \includegraphics[width=0.45\textwidth]{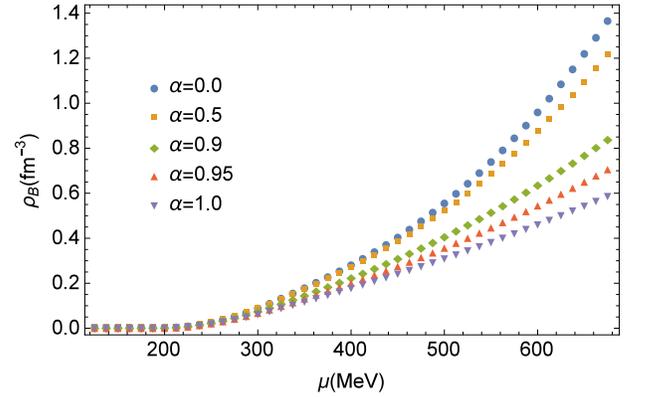}
       \caption{The baryon number density as a function of $\mu$ is presented. When $\mu<M_q$, the quark number density $\rho(\mu)$ is zero. Also $\rho(\mu)$ decreases as $\alpha$ increases and dramatically as $\alpha>0.5$ at large $\mu$.  }\label{fig.mrho}
   \end{center}
\end{figure}
At zero temperature, the quark number density is given by:

\begin{eqnarray}
\rho_{u,d}= 2N_c \int \frac{ d^3p} {(2\pi)^3} \theta(\mu_r-E).
\end{eqnarray}
The $\alpha$-dependence of  effective quark mass and baryon number density are shown in Fig. \ref{fig.mq} and Fig. \ref{fig.mrho} respectively.
The quark mass decreases  as chemical potential increases.  $\alpha$ has an impact on the  effective quark mass. When $\alpha$ is greater than $0.9$, dependence of  quark mass on $\alpha$ is obvious. The quark baryon number density increases  as chemical potential increases. At fixed chemical potential, the quark baryon number density decreases  as $\alpha$ increases.
 Comparing Fig. 2, we can see that at the mean field level the contribution of the effective potential from the Fierz-transformed Lagrangian to the baryon number density is negligible at low chemical potential (for example less than $400$ MeV). Specifically, $\alpha=0$ is equivalent to not considering the contribution of the Fierz-transformed Lagrangian to the effective potential. That is, the contribution of the next-to-leading order term in large $N_c$ expansion is not considered, while the case of  $\alpha \neq 0$ is equivalent to considering the influence of the next-to-leading order term in large $N_c$ expansion on the effective potential.

\section{The quark star structures}
\subsection{The equation of state}
The   EOS is the key to calculating the mass-radius relation and tidal formability $\Lambda$.
The  tidal formability measures the star's quadrupole deformation in response to the companion's perturbing tidal field during the merger of two stars.
%Soft \deleted{equation of state} \added{EOS} cannot support large pulsar mass.

The model-independent  equations of state  of strong interaction matter at finite $\mu$ and zero $T$ are  \cite{zong2008,zong2008a}:

\begin{eqnarray}
P(\mu_u,\mu_d)&=&P(\mu =0) + \sum_{u,d}\int_ 0^{\mu }d\mu \rho  (\mu ),\\
\epsilon(\mu_u,\mu_d)&=&-P(\mu_u,\mu_d)+\sum_{u,d}\mu\rho(\mu).
\end{eqnarray}
Here, $P(\mu =0)$ represents the vacuum pressure  at $\mu=0$. In some works $-P(\mu =0)$ is taken  as a free parameter corresponding to the bag constant in the MIT bag model. In the standard NJL model, it is sometimes  defined at zero chemical potential as the pressure difference between results from Nambu solution and   bare quark propagator:
\begin{eqnarray} \label{eq.pm}
P(\mu=0)=P(M_{N})- P(m),
\end{eqnarray}
where ${M_N}$   denotes the Nambu solution of the quark
gap equation at $\mu=0$ and $m$ is the current quark mass. In this definition, the vacuum pressure   is  $P(\mu=0)=-(129.71$ MeV)$^4$. However, $m$ is not a solution of the gap equation. A consistent definition is to take the difference between two solutions in analogy to the $BCS$ theory. As suggested in Refs. \cite{xu2018,licm,cui2018} the vacuum pressure   in use is:
\begin{eqnarray}
P(\mu=0)=P(M_{N})- P(M_{W}),
\end{eqnarray}
with $M_{W}$ the  quasi--Winger solution of the gap equation.
In this case, the vacuum pressure  is  $P(\mu=0)=-(131.75$ MeV)$^4$.

\begin{figure}
   \begin{center}
      \includegraphics[width=0.45\textwidth]{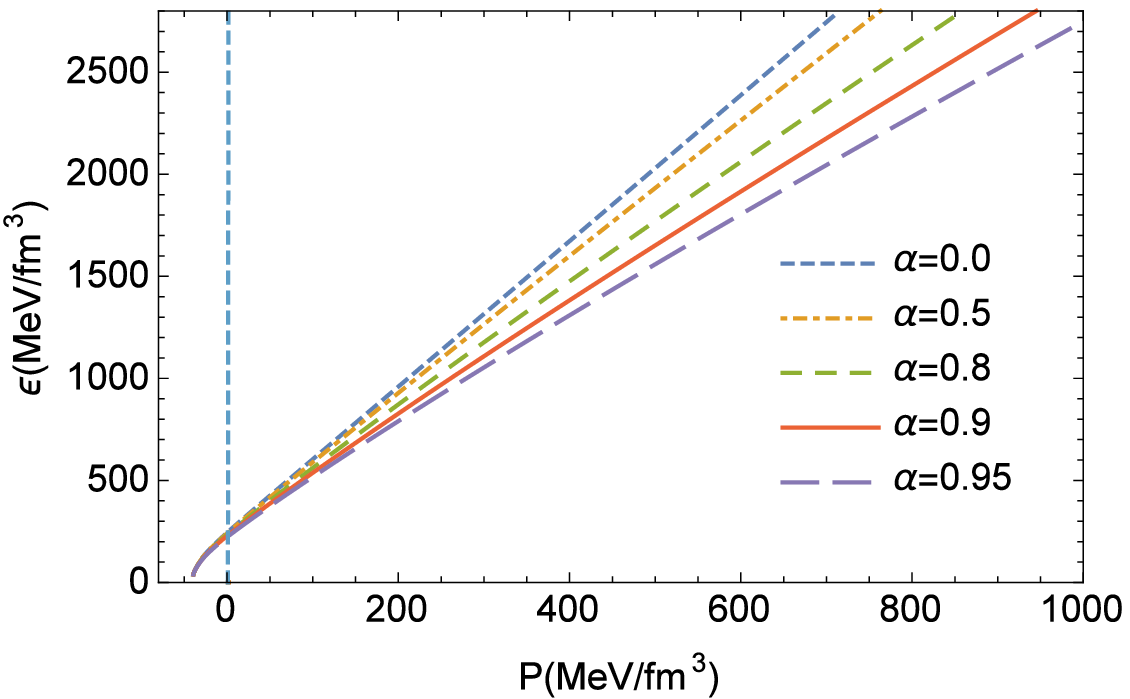}
      \includegraphics[width=0.45\textwidth]{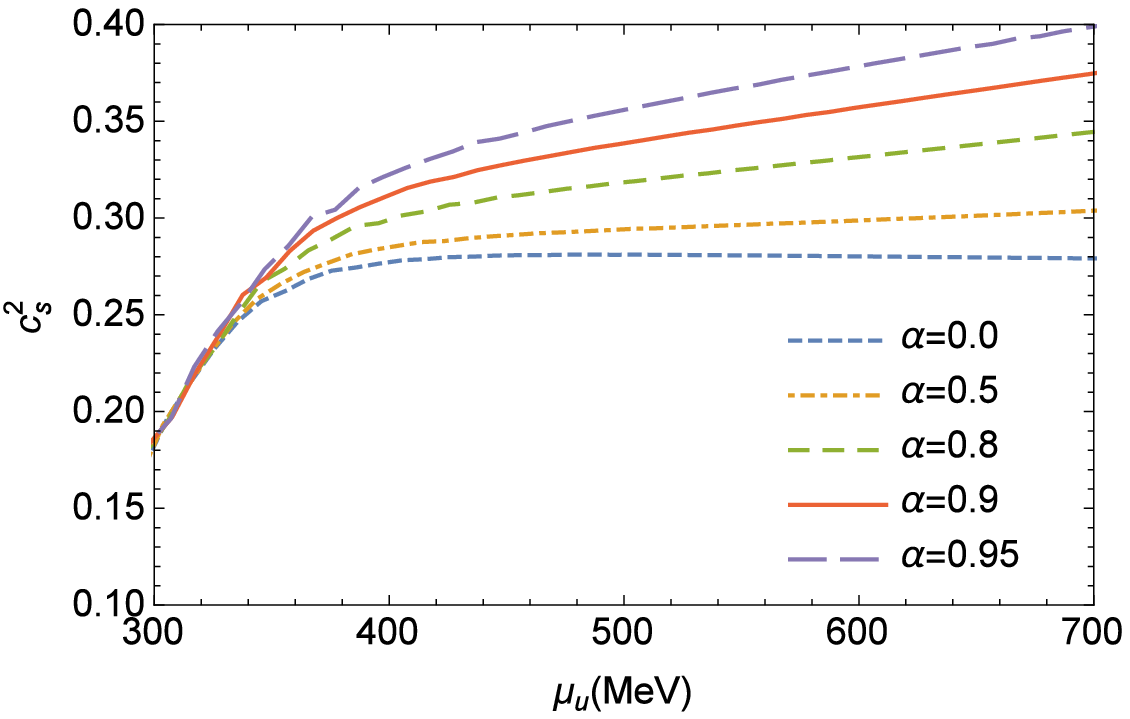}
       \caption{ The EOSs for different $\alpha$'s are presented. The up plane shows energy density $\epsilon$ as function of pressure  $P$. The down plane shows the square of velocity of sound (defined as $\frac{dP}{d\epsilon}$) as function of $u$ quark chemical potential $\mu_u$. Zero  energy density and negative pressure   appear  as $\mu_u<M_q$, and this part of plot is not shown in the figure. Different $\alpha$'s show different results only when $\mu_u$ is greater than $300$ MeV. Both plots show that the EOS for smaller $\alpha$  becomes softer than the one for larger $\alpha$.}\label{fig.peplot}
   \end{center}
\end{figure}

We have only considered pressure from quarks above. For a non-strange quark star, we need electron to keep electric charge neutrality

\begin{equation}
 \frac{2}{3}\rho_u-\frac{1}{3}\rho_d- \rho_e=0, 
\end{equation}
 with $\rho_u$, $\rho_d$, and $\rho_e$ being the number densities of up, down quarks and election respectively. Then the pressure and energy density are
\begin{eqnarray}
P_{tot}&=&P(\mu_u,\mu_d ) +  \frac{\mu_e^4}{12\pi^2},\\
\epsilon_{tot}&=&\epsilon(\mu_u,\mu_d)+  \frac{\mu_e^4}{4\pi^2},
\end{eqnarray}
respectively. Here $\mu_e$ is the electron charge chemical potential and the electron density is given by $\rho_e= \mu_e^3/(3\pi^2)$. We have  to take into account baryon number and electric charge conservation in weak decay $d\leftrightarrow u+e+\bar\nu_e$. Then the chemical potential equilibrium gives $ \mu_d=\mu_u+\mu_e$. Consequently, the   EOS could be obtained. The EOSs with different $\alpha$'s are plotted in  Fig. \ref{fig.peplot}. With a fixed
negative pressure of vacuum, the stiffness of EOS increases along with $\alpha$.

\begin{figure}[h]
   \begin{center}
      \includegraphics[width=0.45\textwidth]{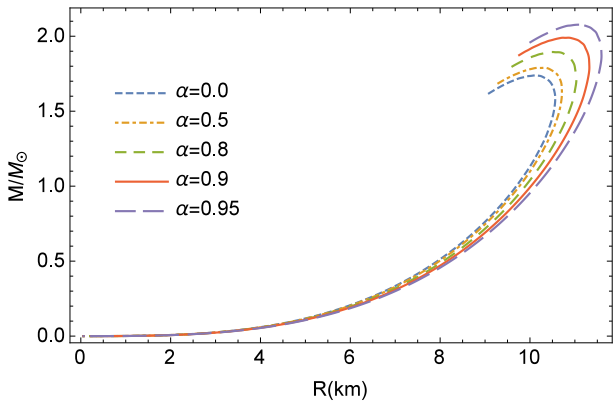}
       \caption{The mass-radius relation of quark star is presented.  The maximum mass
increases with $\alpha$. The maximum radii are less than $12$ km. When $\alpha=0.9$ the maximum mass is close to 2 $M_{\odot}$ with a star radius $10.87$ km. When   $\alpha=0.95$, the maximum mass is larger than $2.05$ $M_\odot$. Only for the case $\alpha \geq 0.5$ The radii of stars with mass of $1.6$ $M_\odot$ are larger than $10.7$ km.}\label{fig.mrplot}
   \end{center}
\end{figure}

We are now ready to investigate   the structure of a quark star   using  the  Tolman-Oppenheimer-Volkoff equations
(in units $G=c=1$ )
\begin{eqnarray}
 \frac {dP\left(r \right)} {dr} &=&-\frac {\left(\epsilon +P \right)\left(M + 4\pi r^3 P \right)} {r \left(r-2M \right)},\\
\frac {dM\left(r \right)} {dr} &=&4\pi r^2 \epsilon,
 \end{eqnarray}
which give the mass-radius relation.
We have calculated quark stars with Eq. (\ref{eq.la})   with different parameters $\alpha$.  The mass--radius relation is presented in Fig.~\ref{fig.mrplot}.

We first notice that as $\alpha>0.9$, the maximum mass (denoted as $M_{TOV}$) can be larger than 2.0 solar masses. This matches the observed pulsars mass of
  PSR J$1614-2230$  $(M=1.928\pm 0.017~ M_{\odot})$ \cite{Fonseca} and  PSR  J$0348+0432$ $(M=2.01\pm 0.04~ M_{\odot})$ \cite{Antoniadis}.
We have also listed the surface energy density $\epsilon_0$  in Tab. \ref{tab.mtov} which satisfy the constrain that $\epsilon_0$ should be larger than $ 2.80\times10^{14}$g/cm$^3 $. Secondly, the upper limit on the radius of a 1.4-solar-mass   star from recent three works are $R \leq 13.76~$km, $R \leq 13.6~$km and $8.9~$km$ \leq R \leq 13.76~$km  respectively  \cite{Fattoyev,Annala,SDe}.   We see in Fig. \ref{fig.mrplot} that the upper limits of radius are all larger than our maximum radii and thus the radii of 1.4-solar-mass star. The lower limit of the radius from reference  \cite{Bauswein2019} on a 1.6-solar-mass neutron star is $10.7~$km. The radii from our parameters $\alpha\geq0.5$   satisfy  this constraint.  If the compact objects are quark stars,  then  in studying QCD matter at high density, Lagrangian without Fierz transformation ($\alpha=0$)   is incomplete at the level of mean field approximation, while Eq.~(\ref{eq.la}) provides a more realistic description.

\begin{table}
\centering
\caption{Properties of $ud$ quark star, including the maximum mass $M_{TOV}$ and the corresponding   radius $R$ and  surface energy density $\epsilon_0$. The radii of 1.6 solar mass star are also listed. It is obvious that the maximum mass increases as $\alpha$ increases.}	\label{tab.mtov}
\begin{tabular}{ccccccccc}
\hline \hline
$\alpha$& &  $M_{TOV} (M_{ \odot })$& & $R$ (km) & & $R_{1.6}$(km)  &       & $\epsilon_0 (10^{14}$g/cm$^3)$\\\hline
0.00& & 1.7376& & 10.17& &10.56 &     & 4.3252 \\
0.50 & & 1.7896& & 10.33& & 10.71&   & 4.2647 \\
0.80 & & 1.8949& & 10.50& & 10.98 &  & 4.1532 \\
0.90& & 1.9903&  & 10.87&  & 11.19 &    & 4.0412 \\
0.95 & &  2.0773& & 11.14& &11.37 &   & 3.9416 \\
 \hline\hline
\end{tabular}
\end{table}

\subsection{The tidal deformability}
During the  merger of two stars, it is reasonable to think about the magnetic breaking during the merger evolution. Love number measures the distortion of the shape of the surface of a star by an external tidal field. The tidal deformability is related to the $l=2$ dimensionless tidal
Love number $k_2$.  In units $G=c=1$, it is
\begin{equation}\label{eq.klam}
  k_2=\frac{3}{2}\Lambda\left(\frac{M}{R}\right)^5.
\end{equation}
 The most recently analysis on the binary neutron star merger GW170817 have found tighter constraints on the component mass to lie between  1.00  and 1.89 $M_\odot$
with $\Lambda$ in (0, 630)  when allowing for large component spins and
 on the component masses to lie between   1.16  and 1.60  $M_\odot$ with $\Lambda = $ $300^{+420}_{-230}$  when the spins are restricted to be within the range observed in Galactic binary neutron stars \cite{Abbott2019}.
 The early restriction on the tidal deformability $\Lambda$ for a $1.4~  M_{\odot} $ is  less than 800~(1400)~ for low~(high)~--spin prior case \cite{Margali,Abbott}.

In matching the interior and exterior solutions across
the star surface, the $l=2$ tidal Love number $k_2$ for the internal solution is given by  \cite{Damour}
\begin{eqnarray}
k_2&=&\frac{8}{5} C ^5   (1 - 2 C)^2 [2 + 2 C (y - 1) - y] \nonumber\\
&& \times \{2 C [6- 3 y+ 3 C (5 y - 8)]\nonumber\\
&& + 4 C^3 [13 - 11 y + C (3 y - 2)+ 2 C^2 (1 + y)] \nonumber\\
&&+    3 (1 - 2 C)^2 [2 - y+ 2 C (y - 1)] ln(1 - 2 C)\}^{-1},\nonumber\\
\end{eqnarray}
where $C=M/R$ defines the compactness of the star and $y$ is related to the metric variable $H$ and surface energy density $\epsilon_0$

\begin{equation} \label{eq.y}
y=\frac{R\beta(R)}{H(R)}-\frac{4\pi R^3 \epsilon_0}{M}.
\end{equation}
 For some neutron star model the surface energy density is zero. But in our NJL model with negative vacuum pressure, the surface energy $\epsilon_0$ is nonzero as shown in Fig. \ref{fig.peplot}.

 The  metric variable $H$ related to the   EOS can be obtained by integrating two differential equations
 \begin{eqnarray} \label{eq.h}
 &&\frac{dH(r)}{dr}=\beta, \\
 \frac{d\beta(r)}{dr} &=&2 gH\{-2 \pi[5\epsilon + 9 P + f (\epsilon +  P)]\nonumber\\
   && + \frac{3}{r^2} +  2 g (\frac{M}{r^2} + 4 \pi r P)^2\}+\nonumber\\
 &&  2g \frac{\beta}{r} [-1 + \frac{M}{r} + 2 \pi r^2 (\epsilon -  P)],
  \end{eqnarray}
where $g=(1-2M/r)^{-1}$ and $f=d\epsilon /dP $. The iteration start from the center at $r=0$ via expansions $H(r) =a_0 r^2 $ and $\beta(r)=2a_0r$ with constant $a_0$.  As can be seen from Eq. (\ref{eq.y}), we only concern the ratio $\beta/H$. So $a_0$ can be arbitrarily chosen   in numerical calculation. The Love number $k_2$ and tidal deformability $\Lambda$ for different $\alpha's$ are calculated and presented in Tab.~\ref{tab.k2}. They  increase as $\alpha$ increase.  For   $1.4~ M_\odot $ star, all the calculated tidal deformabilities lie within a reasonable range, i.e., less than the upper limit of low-spin star and large than the lower limits $\Lambda_{1.4}> 200$ \cite{Bauswein2019}.

\begin{table}

\caption{Properties of our 1.4 solar mass quark star, including the compactness $M/R$, the Love number $k_2$ and the tidal deformability $\Lambda$. Since the radius increases as $\alpha$ increases, the compactness decreases as $\alpha$ increase. And the tidal deformability  increases as $\alpha$ increases. }	\label{tab.k2}
\begin{tabular}{ccccccc}
\hline \hline
$\alpha$& & M/R& &$k_2$& &$\Lambda$\\ \hline
0.00& & 0.1980& &0.1435& & 314.26\\
0.50 & & 0.1962& &  0.1481& & 339.93\\
0.80 & & 0.1925& &  0.1570& &395.06\\
0.90& &0.1895& &0.1645& & 448.98\\
0.95& & 0.1869& & 0.1708& &499.37  \\
 \hline\hline
\end{tabular}
\end{table}

 %%%%%%%%%%%%%%%%%%%%%%%%%%%%%%%%%%%%

\section{summary}

 Using a recently proposed generalised NJL model Eq.~(\ref{eq.la})   with a new parameter $\alpha$   incorporating different interaction channels  \cite{wangf}, we studied the structure of light quark stars in this work.
  We find the EOS of star   gets harder with a larger $\alpha$, namely, with more contribution from the Fierz-transformed  term $L_{Fierz}$.  We use proper-time regularization to treat the ultra-divergence   so  there is no upper limit for the choice of $\alpha$. This improves upon the momentum cutoff regularization scheme used in  \cite{zhaot}.
The    weighting  parameter $\alpha$ is the only free parameter. Other than to   set the vacuum pressure   to be a free parameter or to define the vacuum pressure with current quark as in Eq.~(\ref{eq.pm}),   our vacuum pressure is fixed to be the difference between pressures from Nambu solution and quasi-Wigner solution  \cite{xu2018}. Then the corresponding  bag constant is obtained   to be $B^{1/4}=131.75$ MeV.

From the TOV equations the mass-radius relation and tidal deformability are calculated for different $\alpha$'s.  As $\alpha>0.9$, the 2.0-solar-mass can be yielded, which matches the masses of  PSR J$1614-2230$  $(M=1.928\pm 0.017~ M_{\odot})$ and PSR J$0348+0432$ $(M=2.01\pm 0.04 ~M_{\odot})$.  Our results of  surface energy density are larger than $ 2.80\times10^{14}$g/cm$^3 $ and the radii for a 1.4-solar-mass star satisfy the constraints   $R \leq 13.76~$km  and $R \leq 13.6~$km \cite{Fattoyev,Annala}.    The lower limit  $10.7~$km  of  a 1.6-solar-mass neutron star is  satisfied   for $\alpha \geq 0.5$. This   suggests that the Fierz-transformed Lagrangian must be included in the combined Lagrangian.   We have also calculated the tidal Love number $k_2$ and the tidal deformability $\Lambda$.  The tidal deformability $\Lambda$ calculated for 1.4 solar mass star increases with $\alpha$ and is within the interval ($200<\Lambda<800$) \cite{Bauswein2019} and the interval ($0<\Lambda<630$) for large components spin, and satisfies constraint with $\Lambda=$ 300$^{+420}_{-230}$  when  restricting the magnitude of the component spins from   analysis of $GW170817$ \cite{Abbott2019}.
Admitting the non-strange quark, our improved NJL model therefore provides a consistent explaination to a variety of astronomical observations.

\acknowledgments
This work is supported in part by the National Natural Science Foundation of China (under Grants No. 11475085,  No. 11535005, No. 11690030, and No.11873030, and No. 11905104), the National Major state Basic Research and Development of China (Grant No. 2016YFE0129300).

\end{document}